\documentclass[11pt, epsf]{article}
\usepackage{epsfig}
\def\ga{\mathrel{\raise.3ex\hbox{$>$\kern-.75em\lower1ex\hbox{$\sim$}}}}
\def\la{\mathrel{\raise.3ex\hbox{$<$\kern-.75em\lower1ex\hbox{$\sim$}}}}

\def\I_M{{I_{\scriptscriptstyle M\times M}}}

\begin{document}

\thispagestyle{empty}
\rightline{IP/BBSR/2005-12}
\rightline{\tt hep-th/0507270}

\vskip 2cm \centerline{ \Large \bf  Bouncing Cosmology in Three Dimensions}

\vskip .2cm

\vskip 1.2cm

\centerline{ \bf Anindya Biswas\footnote{Present address: The Institute of 
Mathematical Sciences, C. I. T Campus, Taramani, Chennai 600113, India. 
E-mail: anindyab@imsc.res.in} and 
Sudipta Mukherji\footnote{Electronic address: mukherji@iopb.res.in}}
\vskip 10mm \centerline{ \it Institute of Physics, 
Bhubaneswar-751 005, India} 
\vskip 1.2cm
\vskip 1.2cm
\centerline{\bf Abstract}
\noindent
We consider a dynamical two-brane in a four dimensional
black hole background with scalar hair. At high temperature this black hole 
goes through
a phase transition by radiating away the scalar. The end phase is a 
topological adS-Schwarzschild black hole. We argue here that for a 
sufficiently low temperature, the brane motion in this geometry is 
non-singular. This results in  a universe which passes over from a contracting 
phase to an expanding one without reaching a singularity.

\newpage
\setcounter{footnote}{0}
\noindent

In standard cosmology, the universe usually starts or/and 
ends in a cosmological 
singularity. Since near the singularity the gravitational interaction is 
strong,
classical general relativity breaks down. Consequently, in order to study the 
fate of 
these singularities, one needs to look for a 
still illusive theory of quantum gravity. It is therefore 
of interest to search for a non-singular 
cosmological model by some how evading Penrose-Hawking singularity 
theorem \cite{PH}.
If this is possible, one can at least perform some further computations around 
these cosmological backgrounds while trusting the perturbative theory of 
gravity.

In the context of brane-world models, where the brane moves in 
a higher dimensional 
bulk, we often get exotic cosmologies on the brane 
(see \cite{RM,KK} for example). 
Since, in principle, 
in this scenario, bulk geometry can contribute a negative energy density on
the brane \cite{DNV}, it may be possible to construct some non-singular 
cosmological models. Indeed, some models of this nature were constructed in
\cite{MP,BMP,OM,BQRTZ}. In particular, in \cite{MP}, 
a non-singular cosmology was 
found by considering a three-brane in an electrically charged adS 
black hole background \cite{BV}. Here, the brane contracts to a finite size 
before  expanding again. This happens because the background charge 
introduces a negative 
energy density on the brane world volume; hence preventing it to fall into the
singularity. However, the bounce occurs inside the outer horizon of 
the black hole. As a result, regardless of the initial energy of the brane, 
the 
bulk space time collapses due to the back reaction at the bounce \cite{HM}.\\
 
In our effort to search for other examples of non-singular cosmology, in this 
Letter, we study another model of bounce in three dimensions. As we will show 
below, in certain 
region of the parameter space of the bulk geometry, the brane makes a 
smooth transition
from a contracting phase to an expanding phase without ever reaching to a 
singularity. 
The model of our interest consists of black holes in four dimensional gravity 
with self-interacting  scalar field 
and a negative cosmological constant. The solutions
are of topology ${\bf R}^2 \times \Sigma$ where $\Sigma$ represents a two 
dimensional
manifold with a constant negative curvature. These holes
are parametrised by a single parameter associated with the mass of the black 
holes. Interestingly enough, the mass of the black hole here can
take negative values without producing a naked singularity. We will take an 
advantage of this fact to construct a bouncing brane cosmology by 
considering a dynamical 
two-brane in this geometry. In the following, we describe the 
model in some detail and, subsequently, analyse the resulting cosmology.

Let us start by considering the action
\begin{equation}
S = \int d^4x\sqrt{-g}\Big[{{R + 6/l^2}\over{16 \pi}} - {1\over 2}\partial_\mu\phi 
\partial^\mu\phi + 
{3\over{4\pi l^2}}{\rm sinh}^2({\sqrt{4\pi}\phi\over{\sqrt 3}})\Big].
\label{act}
\end{equation}
Around the global maximum ($\phi =0$), the scalar satisfies the 
Breitenlohner-Friedman 
bound \cite{BF}
ensuring perturbative stability of the adS space-time. A black hole solution of
this action is given by \cite{MTZ}
\begin{equation}
ds^2 = {r(r + 2 m)\over{(r + m)^2}}\Big[-\Big({r^2\over{l^2}} -
\big(1 + {m\over r}\big)^2\Big)dt^2 
+ \Big({r^2\over{l^2}} -\big(1 + 
{m\over r}\big)^2\Big)^{-1}dr^2 + r^2 d\sigma^2\Big],
\label{met}
\end{equation}
with the scalar
\begin{equation}
\phi = {\sqrt{3\over{4\pi}}}{\rm Arctanh} \Big({m\over{r + m}}\Big).
\label{bulks}
\end{equation}
In (\ref{met}), $d\sigma^2$ represents an element of a  two dimensional 
manifold
with a negative constant curvature ($\Sigma$). The mass of the black hole is 
given
by 
\begin{equation}
M = {\sigma\over{4\pi}}m,
\end{equation}
where $\sigma$ is the area on $\Sigma$. For $m >0$, the solution has 
a singularity at
$r =0$. This singularity is hidden behind a horizon. 
For $m < -l/4$, the metric has a naked singularity.
However, for $-l/4 < m < 0$, the singularity at $r= - 2m$ is surrounded by 
three 
horizons with the outermost horizon is at
\begin{equation}
r_+ = {l\over 2}\Big(1 + {\sqrt{1 + {4m\over l}}}\Big).
\label{hor}
\end{equation}
Note that the origin of the radial coordinate here should be taken at
$r = -2m$. 
The temperature of the
black hole is given by 
\begin{equation}
T = {1\over{2\pi l}}\Big({2 r_+\over l} -1).
\end{equation}
The entropy is proportional to the horizon area and is
\begin{equation}
S = {\sigma\over {4}} {{r_+}^3 (r_+ + 2 m)\over{(r_+ + m)^2}}.
\label{centro}
\end{equation}

One of the interesting properties of this black hole is that it undergoes 
a phase transition. This happens as we increase the temperature above a 
critical temperature given by $T_c = {1/{2\pi l}}$. 
At this point $m \rightarrow 0$ and $r_+ \rightarrow l$. For temperature above 
$T = T_c$, the black hole radiates away its scalar hair
and decays to a topological Schwarzschild-adS black hole with metric
\begin{equation}
ds^2 = -\Big({r^2\over{l^2}} -1 - {2m\over r}\Big)dt^2 + 
\Big({r^2\over{l^2}} -1 - {2m\over r}\Big)^{-1} dr^2 + r^2 d\sigma^2,
\label{sch}
\end{equation}
with temperature
\begin{equation}
T = {1\over{4\pi}}\Big({3 r_+\over l} -{1\over r_+}\Big),
\end{equation}
and entropy
\begin{equation}
S = {\sigma r_+^2\over 4}.
\label{entro}
\end{equation}
Here $r_+$ is the horizon location of (\ref{hor}) and should not be
confused with the $r_+$ appearing in (\ref{met}). 
Note that both the metric (\ref{met}) and (\ref{sch}) have same asymptotic 
geometry.
Furthermore, as can be seen from (\ref{centro}) and (\ref{entro}),
the entropy changes continuously around the transition point.
It is also possible to define an order parameter which changes continuously to 
zero as we increase the temperature above $T=T_c$\cite{MTZ}. 

We would now like to consider a two-brane with three dimensional 
world volume moving in this 
background. We will assume that the brane is much lighter than the 
bulk. Consequently, any back reaction to the background geometry will be 
neglected. 
The world volume action of the brane is
\begin{equation}
S_b = -\int d^3x {\sqrt{-\gamma}}{\hat V(\phi)},
\label{braneac}
\end{equation}
where $\gamma_{ij}$ is the induced metric on the brane and $\hat V(\phi)$ is 
the potential on the brane which we will determine shortly.\\ 

Let us assume that the
bulk metric has a general form
\begin{equation}
ds^2 = -A(r)dt^2 + B(r) dr^2 + R^2(r) d\sigma^2.
\label{bmet}
\end{equation}
If we write the induced brane metric as
\begin{equation}
ds_b^2 = \gamma_{ij} dx^i dx^j = - d\tau^2 + R^2(\tau) d\sigma^2,
\label{indu}
\end{equation}
then it follows immediately from (\ref{bmet}) and (\ref{indu}) that
\begin{equation}
\Big({\partial \tau\over{\partial t}}\Big)^2 = {A\over{1 + B \Big({\partial 
r\over{\partial \tau}}\Big)^2}}.
\label{ir}
\end{equation}
Finally, using (\ref{ir}) in the Israel junction condition (as 
discussed, for 
example, in \cite{CR}) we get
\begin{equation}
{1\over 2}\Big({dR\over{d\tau}}\Big)^2 + F(R) = 0,
\label{isr}
\end{equation}
where $F(R)$ is given by
\begin{equation}
F(R) = {1\over{2B}}\Big({dR\over{dr}}\Big)^2 - {\hat V^2R^2\over{32}}.
\end{equation}
Above equation can  also be expressed as
\begin{equation}
{1\over 2} \Big({dr\over {d\tau}}\Big)^2 + U(r) = 0,
\label{enereq}
\end{equation}
where the potential term 
\begin{equation}
U(r) = {1\over 2 B} - {\hat V^2 R^2\over{32 {R^\prime}^2}}.
\label{effctv}
\end{equation}
In writing (\ref{enereq}), we used the fact that $R^\prime = 
dR/dr$ is nonzero everywhere in the region $r > -2m$ and it 
goes to a constant value at large $r$.
Eqn. (\ref{enereq}) can be thought of as a 
Hamiltonian constraint for a zero energy classical particle.

We now turn our attention to the brane potential $\hat V(\phi)$.
Firstly, $\hat V$ can not be chosen arbitrarily. It is fixed by
the boundary condition on the bulk scalar field at the brane.
Following \cite{CR}, the boundary condition on this scalar 
at the wall can be written as 
\begin{equation}
\{n.\partial \phi\} = {d\hat V\over{d \phi}},
\end{equation}
where $n$ is the unit normal at the boundary pointing towards the
bulk and the curly brackets denote summation over both sides of the
wall. The above equation can be simplified \cite{CR} using (\ref{isr})
and (\ref{met}) to get
\begin{equation}
{d\phi\over dR}= -{2\over R} {d\over {d\phi}}\log\hat V.
\label{scalarbc}
\end{equation}
For the scalar profile (\ref{bulks}), one can solve (\ref{scalarbc})
explicitly with the result
\begin{equation}
\hat V= \lambda\Bigg[{(3m^2+3mr+r^2)^2\over {r(r+2m)^3}}\Bigg]
^{1\over{16\pi}},
\label{pot1}
\end{equation}
where, $\lambda$ is an integration constant and related to the
brane tension. This can be seen by taking $r 
\rightarrow \infty$ 
limit of (\ref{pot1}) and comparing it with the boundary action
$S_b$ given in eqn. (\ref{braneac}). Though not very illuminating, 
$\hat V$ can indeed be expressed as a function of the scalar $\phi$ only. 
It is given by
\begin{equation}
\hat V(\tilde\phi) = \lambda \Bigg[ {{\rm coth}^2 \tilde \phi + {\rm coth}~\tilde \phi + 1\over{
{\rm cosech} ~\tilde \phi ~({\rm coth} ~\tilde \phi + 1)}}\Bigg]^{1\over{8\pi}},
\label{pot2}
\end{equation}
where $\tilde\phi = {\sqrt{4 \pi/3}}\phi$. Note that the explicit
dependence of the black hole mass term $m$ has disappeared in (\ref{pot2}).
We therefore conclude that even if in (\ref{braneac}), we have introduced $\hat V(\phi)$ in
somewhat adhoc manner, the potential gets determined (upto a constant $\lambda$)
via the bulk scalar boundary condition on the brane.

For our metric (\ref{met}), $U(r)$ in (\ref{effctv}) can  
be explicitly computed. It is given by
\begin{equation}
U(r) = {1\over 2} (r +m)^2 \Bigg[{{r^2\over l^2} - (1+ {m\over r})^2
\over{r(r + 
2m)}} - {{\lambda^2(r^2+3mr+3m^2)^{1\over4\pi}\over 16(r(r+2m)^3)^{1\over
  8\pi}}}{r^2(r + 2m)^2 \over{(r^2 + 3mr +3m^2)^2}}\Bigg].
\label{pot}
\end{equation}
Following our earlier discussion, we note that for higher temperature, 
the black hole goes through a phase transition. Hence, equation (\ref{enereq}) 
describes the correct behaviour of the brane universe only for $T < T_c$.
For temperature $T >T_c$, the bulk is a topological Schwarzschild-adS black 
hole. 
Therefore, the effective potential $U(r)$ modifies to 
\begin{equation}
\tilde U(r) = {1\over 2}\Big[- 1 + \Big({1\over l^2} - 
{\lambda^2\over{16}}\Big)r^2 - 
{2 m\over r}\Big].
\label{tlarge}
\end{equation}
This follows again from the Israel junction condition when the bulk 
is given by metric (\ref{sch}).

We would now be interested in solving the eqn. (\ref{enereq}) 
for $T<T_c$ where 
the black hole mass is negative and  in the region $-l/4 < m < 0$.
Since, $R^\prime$ is a smooth non-zero function in the range $r >-2 m$, the 
cosmological nature of the brane can be inferred from (\ref{enereq}) and 
(\ref{pot}). First of all, this behaviour crucially depends on the 
signature of $\lambda^2 - 16/l^2$. We will mainly be interested in the case 
when 
this quantity is positive. 

As it is not possible to solve (\ref{enereq}) analytically 
in a compact form, we will 
have to resort to solving it by numerical means. However, before we do so, 
we will try to understand the qualitative nature of 
the solution from the potential (\ref{pot}) itself.
Since for large $r$, 
\begin{equation}
U(r) \sim \Big({1\over l^2} - {\lambda^2\over{16}}\Big) r^2,
\end{equation}
for $\lambda^2 > 16/l^2$, we get the deSitter like solutions. 
On the other hand, near the singularity at $r = -2 m$
\begin{equation}
U (r) \sim m\Big({1\over {16}} - {m^2 \over l^2}\Big) (r + 2 m)^{-1}.
\end{equation}
which is independent of the tension of the brane.
One of the most interesting property of the potential is that
for certain range of the mass of the black hole
there is a turning point.  This can be seen clearly in fig.1. 
\begin{figure}[ht]
\epsfxsize=8cm
\centerline{\epsfbox{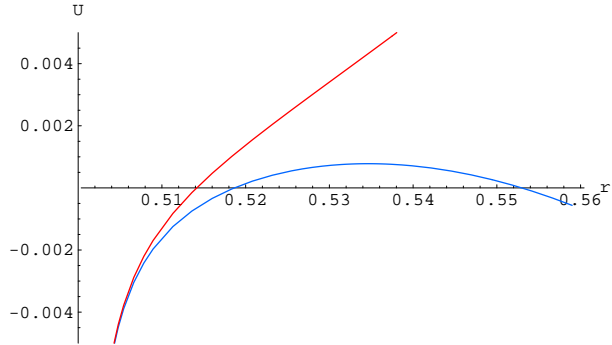}}
\caption{{\small{The bottom line is the potential $U(r)$. The largest $r$ for 
which $U(r) = 0$ is the location of the bounce. The largest value of $r$ at 
which the top line crosses the horizontal axis corresponds to the outer 
horizon. These plots are for $l =1, \lambda = 4.1$ and $m = -.2498$. }}} 
\end{figure}
This point corresponds to the largest solution $r_b$
of the equation $U(r) =0$. The expression of 
$U(r)$ is given in (\ref{pot}).
As this point is outside the outer most horizon
of the black hole, the brane will bounce back even before reaching the 
horizon. This will result in a bouncing cosmology where the universe reaches a 
minimum size $R(r_b)$ and then expands again.
It is straightforward to integrate (\ref{enereq}) numerically. 
The result is shown in fig.2. 
\begin{figure}[ht]
\epsfxsize=8cm
\centerline{\epsfbox{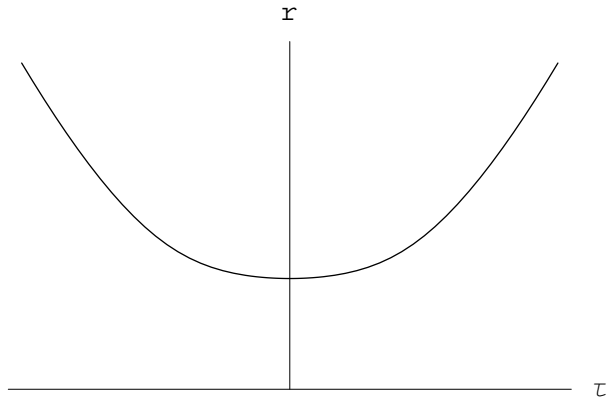}}
\caption{{\small{Solution of eqn.(\ref{enereq}). Time is chosen such
that  the bounce is at $\tau =0$.}}}
\end{figure}

In case of $\lambda^2$ sufficiently less than $16/l^2$, all the solutions are singular.
In this case, the brane can start at a finite 
radius and can fall into the black hole singularity, resulting in a 
singular brane universe. Typical nature of the potential is shown in 
fig.3.
\begin{figure}[ht]
\epsfxsize=8cm
\centerline{\epsfbox{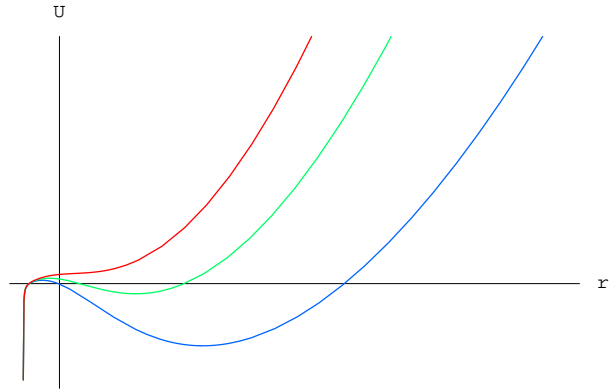}}
\caption{{\small{$U(r)$ for $\lambda^2 < 16/l^2$. In 
$\lambda$ is sufficiently less than $16/l^2$ brane starts at a 
finite radius and falls into the black hole singularity. Plots are for
different values of $\lambda^2$.}}}
\end{figure}

For fixed $l$ and $\lambda$, if we increase the temperature of the 
black hole beyond $T_c$, due to the phase transition, the effective 
potential felt by the brane is given by eqn.(\ref{tlarge}). This always
leads to a singular cosmology as the brane collapses into the 
adS-Schwarzschild singularity (Discussions of this can be found, for example,  in
\cite{PS}). 

To conclude, we therefore have the following picture. Consider a 
brane with fixed energy density (larger than the scale set by the 
adS curvature) in the asymptotic region of a 
topological adS-Schwarzschild geometry. As we decrease
the temperature of the black hole below a critical value, the 
bulk goes through a phase transition. In this phase, the 
geometry is that of a black hole with scalar hair. Consequently, the 
singular cosmology of the brane universe at high temperature
modifies to a bouncing cosmology at low temperature. Note that, unlike 
some other models,
this bounce takes place even before the brane reaches the horizon
of the bulk. Before we end,
we would like to point out that our conclusion is based on a 
completely classical 
analysis. This picture might get modified in several way. 
Firstly, instead 
of an empty brane, one expects the brane to carry additional matter. This will 
change the behaviour of the brane universe in a significant way. Secondly, 
various instabilitis may creep in when the model is treated semi-classically.
We leave all these issues for future considerations.

\noindent{\bf{Acknowledgments:}} 
We wish to thank Tanay Kumar Dey for discussions.


\begin{thebibliography}{99}


\bibitem{PH} S.W. Hawking and G.F.R. Ellis, `The large scale structure
of space time', Cambridge University Press, Cambridge, England (1973).

\bibitem{RM} R. Maartens, Living Rev. Rel. 7 (2004) 7,
gr-qc/0312059. 

\bibitem{KK} A. Kehagias and E. Kiritsis, JHEP 9911 (1999) 022,
hep-th/9910174.

\bibitem{DNV} D.N. Vollick, Gen. Rel. Grav. 34 (2002) 1,
hep-th/0004064.

\bibitem{MP} S. Mukherji and M. Peloso, Phys. Lett. B547 (2002) 297,
hep-th/0205180.

\bibitem{BMP} A. Biswas, S. Mukherji and S.S. Pal, Int. J. Mod. Phys. A19
(2004) 557, hep-th/0301144.

\bibitem{OM} N. Okuyama and K. Maeda, Phys. Rev. D70 (2004) 064030,
hep-th/0405077.

\bibitem{BQRTZ} C.P. Burgess, F. Quevedo, R. Rabadan, G. Tasinato, I. Zavala, 
JCAP 0402 (2004) 008, hep-th/0310122.

\bibitem{BV} C. Bercelo and M. Visser, Phys. Lett. B482, (2000) 183,
hep-th/0004056.

\bibitem{HM} J.L. Hovdebo and R.C. Myers, JCAP 0311 (2003) 012, 
hep-th/0308088.

\bibitem{BF} P. Breitenlohner and D.Z. Freedman, Phys. Lett. B115 (1982) 
197; Annals. Phys. 144 (1982) 249.

\bibitem{MTZ} C. Martinez, R. Troncoso and J. Zanelli, Phys. Rev. D70 (2004)
084035, hep-th/0406111.

\bibitem{CR} H.A. Chamblin and H.S. Reall, Nucl. Phys. B562 (1999) 133,
hep-th/9903225.

\bibitem{PS} A.C. Petkou and G. Siopsis, JHEP 0202 (2002) 
045, hep-th/0111085.


\end{thebibliography}
\end{document}